\documentclass{ws-ijmpd}

\begin{document}

\markboth{Joseph Sultana, Demosthenes Kazanas} {The Problem of
Inertia in Friedmann Universes}

\catchline{}{}{}{}{}
%
%%%%%%%%%%%%%%%%%%%%%%%%%%%%%%%%%%%%%%%%%%%%%%%%%%%%%%%%%%%%%%%%%%%%

\title{THE PROBLEM OF INERTIA IN FRIEDMANN UNIVERSES}

\author{JOSEPH SULTANA}

\address{Department of Mathematics, Faculty of Science, University of Malta\\
Msida MSD2080, Malta\\
joseph.sultana@um.edu.mt}

\author{DEMOSTHENES KAZANAS}

\address{Astrophysics Science Division, NASA/Goddard Space Flight Center\\
Greenbelt, Maryland 20771, USA\\
demos.kazanas@nasa.gov}

\maketitle

\begin{history}
\received{Day Month Year}
\revised{Day Month Year}
\end{history}

\begin{abstract}
In this paper we study the origin of inertia in a curved spacetime,
particularly the spatially flat, open and closed Friedmann
universes. This is done using Sciama's law of inertial induction,
which is based on Mach's principle, and expresses the analogy
between the retarded far fields of electrodynamics and those of
gravitation.  After obtaining covariant expressions for
electromagnetic fields due to an accelerating point charge in
Friedmann models, we adopt Sciama's law to obtain the inertial force
on an accelerating mass $m$ by integrating over the contributions
from all the matter in the universe. The resulting inertial force
has the form $F = -kma$, where $k < 1 $ depends on the choice of the
cosmological parameters such as $\Omega_{M},\ \Omega_{\Lambda}, $
and $\Omega_{R}$ and is also red-shift dependent.

\keywords{Inertia; Friedmann Universes; Mach's Principle.}
\end{abstract}

%\ccode{PACS numbers: 04.20.Cv, 98.80.Jk}

\section{Introduction}

The concept of inertia has been one of the most debated topics of
classical physics, starting with Newton's ideas of absolute space
and that of inertia as an intrinsic property of matter devoid of any
external influence. The notion of absolute space was criticized by
Leibniz and later Bishop Berkeley who claimed that it is
metaphysical. They were followed by Ernst Mach \cite{mach} who in
1872 rejected the existence of absolute space in favour of relative
motion with respect to a ``fixed'' frame provided by the matter
distribution in the universe, and claimed that it is the
acceleration relative to this frame that determines the inertial
properties of matter. This is the essence of Mach's Principle, a
term coined by Albert Einstein in 1918, which says that ``the
inertial force which acts on an accelerating object is due to its
interaction with all matter present in the rest of the universe''.
There are indeed several interpretations of Mach's principle and
arguments on whether the general relativity theory is Machian or
not.\cite{barbour}\cdash\cite{sachs} Besides Mach's principle there
are other approaches towards the explanation for the orgin of
inertia which describe inertia as a local instead of a global
phenomenon. Among these one can mention the suggestion that inertial
forces result from the interaction of matter with electromagnetic
fluctuations of the zero point field, \cite{haisch} (which is
assumed to be homogeneous and isotropic, a property likely also
related to the large scale distribution of matter in the universe
and thus to the same frame envisioned by Mach) or the attribution of
inertia to the interaction of a particle with its own field, namely
the self-force.\cite{martins}

The implementation of Mach's principle as a quantitative law was
done by D.W. Sciama \cite{sciama} in 1953, by applying the formalism
of electrodynamics to gravitation. According to Sciama the
gravitational field of a moving particle in the universe can be
expressed in terms of a ``gravoelectric'' part $\mathbf{E}$ and a
``gravomagnetic'' part $\mathbf{B}$. These can be written in terms
of the retarded scalar and vector potentials $\phi$ and $\mathbf{A}$
by
\begin{eqnarray}
\begin{array}{ccl}
\mathbf{E} &=& - \mathrm{grad}\phi -
\left(\frac{1}{c}\right)\frac{\partial\mathbf{A}}{\partial t}\\[3mm]
\mathbf{B} &=& \mathrm{curl}\mathbf{A},
\end{array} \label{sciama1}
\end{eqnarray}
where $\phi$ is taken to be the Newtonian potential and $\mathbf{A}
= (\phi/c)\mathbf{v}(t)$; $\mathbf{v(t)}$ being the velocity of the
particle.

According to this law, the gravitational force between two masses
$m_{1}$ and $m_{2}$ with relative acceleration $a$ can be expressed
by
\begin{equation}
F = \frac{Gm_{1}m_{2}}{r^2} + \frac{Gm_{1}m_{2}a}{c^2 r},
\label{sciama2}
\end{equation}
in analogy with the Lorentz force, where the acceleration dependent
term is referred to as ``inertial induction'' and is the result of
acceleration in fields that obey retardation. Since the universe can
be taken to be isotropic on the large scale, the $1/r^2$ term will
cancel when integrating over all masses in the universe to find the
resulting inertial force on an accelerating test particle of mass
$m$. This force is therefore given by \cite{berry}
\begin{equation}
F = \sum_{\mbox{all}\,M}\frac{GmMa}{c^2 r} =
\frac{Gma\tilde{\rho}}{c^2}\int\!\int\!\int_{R_{U}}\frac{1}{r}\mathbf{\mathrm{d}r},
\label{berry1}
\end{equation}
where the region $R_{U}$ is taken by Berry (see
Ref.~\refcite{berry}) to be the observable universe having density
$\tilde{\rho}$ given by that part of the universe which is contained
within the Hubble sphere with radius $c/H$, representing the
distance where galaxies are receding with the speed of light. Hence
\begin{equation}
F = \frac{4\pi Gma\tilde{\rho}}{c^2}\int^{c/H}_{0}r\,\mathrm{d}r =
\left(\frac{2\pi G\tilde{\rho}}{H^2}\right)ma. \label{berry2}
\end{equation}
Substituting values for $\tilde{\rho}$ and $H$ in the coefficient of
$ma$ in equation (\ref{berry2}), Berry obtained a value of
$\frac{1}{25}$ instead of unity. He explained that this discrepancy
results from the uncertainty in $\tilde{\rho}$ which does not
include dark matter and the increase in masses of distant galaxies
by the relativistic factor $(1 - v^2/c^2)^{-1/2}$. \newline An
attempt to generalize Mach's principle to curved spaces,
particularly the Friedmann universes, was done by R. L.
Signore,\cite{signore1}\cdash\cite{signore2} by using Sciama's law
with a potential
\begin{equation}
\phi = \int\left(\frac{\ddot{R}}{R}\right) r\,\mathrm{d}r,
\label{signore}
\end{equation}
where $R(t)$ is the scale factor, instead of the above Newtonian
potential. Moreover instead of limiting the volume of the universe
to the Hubble sphere, the above integration is extended to the
causal sphere of the accelerating test particle represented by the
particle horizon of the particle. In the case of the spatially flat
Einstein-de Sitter model this potential takes the simple value of
$-c^2$, and then (\ref{sciama1}) gives the result $F = ma$, in
accordance with the equivalence principle.  In the case of a
spatially curved Friedman model the calculation leads to $F =
\lambda ma$ where $\lambda >1$ is time dependent. This means that
either the time dependent inertial mass is larger than the
gravitational mass, or else, the gravitational constant $G$ changes
with time to keep the ratio of the gravitational to inertial forces
constant. However Signore's work is based on two assumptions which
in our opinion limit the validity of the obtained results. The first
one involves the integration in (\ref{signore}), which as in Berry's
calculation (\ref{berry2}), assumes that the density of the universe
$\rho$ is constant over its entire history. Moreover Signore claimed
that in a curved space the field acting on an accelerating test
particle is still given by the non-covariant flat space expressions
in (\ref{sciama1}).

In this article we relax these two assumptions and use Sciama's
approach to find the inertial force on an accelerating test particle
arising from the inertial induction of all matter in the universe.
As done by Signore we consider the particle horizon instead of the
Hubble sphere to mark the region of the universe that influences our
local dynamics.\cite{davis} This represents the distance to the
farthest objects that can be observed at any particular time, and
the redshift becomes infinite for emitters on the particle
horizon.\footnote{We can receive photons emitted by receding
emitters beyond the Hubble sphere ($v_{\mbox{rec}} > c$), because
the Hubble sphere expands and overtakes these photons which end up
inside the Hubble sphere where $v_{\mbox{rec}} < c$.} However unlike
previous calculations by Berry and Signore we use a variable denisty
$\rho$ when integrating over the entire universe. Moreover instead
of the expressions for the retarded potentials and fields in
(\ref{sciama1}) which are valid in flat space, we use general
covariant formulations valid in a curved spacetime.  The inertial
force on an accelerating particle for the different types of
Friedmann models is obtained in a form which is covariant with
respect to coordinate transformations in the 3-space of these
models. In section 2, we derive covariant expressions for the
retarded potentials and fields for an accelerating test particle in
the three cases of flat, open and closed Friedmann models. Then in
section 3 we obtain expressions for the inertial forces and the
results are summerized and discussed in section 4.

\section{Covariant Fields}
The study of electromagnetism in a general curved spacetime was done
by DeWitt and Brehme.\cite{dewitt} Later Hobbs\cite{hobbs} obtained
a formal solution for the electromagnetic potentials\footnote{Greek
indices take values from 0 to 3; latin indices represent the spatial
components 1 to 3.} $A^{\mu} = (\phi, \mathbf{A})$ in the case of a
conformally flat spacetime, and showed that in this case, the
electromagnetic signals propagate only on the light cone without
scattering off the Riemann tensor or radiation tail as in general
curved spacetimes, thereby simplifying the solutions.

Let us write the metric for the Friedmann universes in the form
\begin{equation}
ds^2 = - dt^2 + R^2(t)[d\rho^2 +
f_{k}^2(\rho)d\Omega^2],\label{FRW1}
\end{equation}
where
\begin{equation}
f_{k}(\rho) = \Bigg\{\begin{array}{l}
\sin\rho,\quad \mbox{closed}\ (k=1)\\
\rho,\quad \mbox{flat}\ (k=0)\\
\sinh\rho,\quad \mbox{open}\ (k=-1)\label{f}
\end{array}
\end{equation}
$R(t)$ is the scale factor obtained by solving the Friedmann
equations and $d\Omega^2 = d\theta^2 + \sin^2\theta d\phi^2$. Then
the potentials due to an accelerating point charge $q$ moving along
the path $\mathbf{x}' = \mathbf{z}(t')$, where $(t', \mathbf{x}')$
represent the retarded position, are given by\cite{peters}
\begin{eqnarray}
\begin{array}{ccl}
A_{0}(\mathbf{x}, t) & = & \displaystyle\frac{q F(\Psi)}{R(t)\sigma}, \\[2mm]
A_{k}(\mathbf{x}, t) & = & -\displaystyle\frac{q[G(\Psi)d_{km'} +
H(\Psi)n_{k}n_{m'}]v^{m'}(t')}{R(t)\sigma},\label{potentials}
\end{array}
\end{eqnarray}
where
\begin{eqnarray}
\begin{array}{ccrl}
F(\Psi) & = & \cot\Psi, & \quad k = 1,\\[1mm]
        & = & 1/\Psi,\ & \quad k = 0,\\[1mm]
        & = & \coth\Psi, & \quad k =-1,\\[1mm]
G(\Psi) & = & \csc\Psi, & \quad k = 1, \\[1mm]
        & = & 1/\Psi, & \quad k = 0, \\[1mm]
        & = & \mbox{csch}\Psi, & \quad k =-1,\\[1mm]
H(\Psi) & = & \tan\frac{1}{2}\Psi, & \quad k = 1,\\[1mm]
        & = & 0, & \quad k = 0, \\[1mm]
        & = & -\tanh\frac{1}{2}\Psi, & \quad k =-1,\label{FGH}
\end{array}
\end{eqnarray}
and
\begin{equation}
n_{k} = R(t)\Psi_{;k};\quad n_{k'} = R(t')\Psi_{;k'},
\label{unitvectors}
\end{equation}
\begin{equation}
d_{km'} = R(t)R(t')g_{km'},\label{ppropagator}
\end{equation}
\begin{equation}
\sigma = 1 + n_{k'}v^{k'}; \quad v^{k'} =
dz^{k}(t')/dt'.\label{sigma}
\end{equation}
%\newpage
The function $\Psi(\mathbf{x}, \mathbf{x'})$ is the bi-scalar
distance\footnote{In general two-point tensors, or bi-tensors, are
used in the description of physical processes in which a cause at a
point $P'$ brings about an effect at a point $P$. Their indices (or
arguments) refer to the points $P'$ or $P$ and are written with or
without a prime over the index. For more details on bi-tensors see
Synge (1960) in Ref.~\refcite{synge}.} in 3-space measured along a
geodesic joining the points $\mathbf{x}$ and $\mathbf{x'}$. This can
be expressed as $L(\mathbf{x}, \mathbf{x'}, t)/R(t)$, where $L$ is
the distance between $\mathbf{x}$ and $\mathbf{x'}$ at time t. The
vectors $n_{k}$ and $n_{k'}$ in (\ref{unitvectors}) are unit tangent
vectors at $\mathbf{x}$ and $\mathbf{x'}$ to the geodesic joining
these two points, and the bi-vector parallel propagator\cite{synge}
$d_{km'}$ gives a relation between the components of a vector
$A^{i}$ at $\mathbf{x}$ and the components of the same vector
parallel transported along the geodesic joining $\mathbf{x}$ and
$\mathbf{x'}$,
\begin{equation}
A_{k} = d_{km'}A^{m'},\quad A_{m'} = d_{m'k}A^{k}.
\end{equation}
The components of the electromagnetic field tensor $F_{\mu\nu} =
A_{\nu;\mu} - A_{\mu;\nu}$ are given by\cite{peters}
\begin{eqnarray}
%\begin{array}{lll}
F_{0k}(\mathbf{x},t) & = &
\textstyle\frac{1}{R^{2}(t)}\left\{\frac{G(\Psi)F(\Psi)}{R(t')}(d_{km'}
+ n_{k}n_{m'})J^{m'} + \frac{d}{d
s}\left[\frac{G(\Psi)}{R(t')}(d_{km'} +
n_{k}n_{m'})J^{m'}\right]\right. \nonumber\\[3mm]
& &  + \textstyle\left.\frac{G^{2}(\Psi)}{R(t')}(n_{k}n_{m'}J^{m'} -
n_{k}J^{0'})\right\},\label{F0k}
%\end{array}
\end{eqnarray}
\vspace{1mm}
\begin{eqnarray}
F_{km}(\mathbf{x},t) & = &
\textstyle\frac{1}{R^2(t)}\left\{\frac{G(\Psi)
F(\Psi)}{R(t')}(d_{kl'}n_{m} - d_{ml'}n_{k})J^{l'}\right.\nonumber \\
& &  \hspace{42mm}
+\textstyle\left.\frac{d}{ds}\left[\frac{G(\Psi)}{R(t')}(d_{kl'}n_{m}
- d_{ml'}n_{k})J^{l'}\right]\right\},\label{Fkm}
\end{eqnarray}
where
\begin{equation}
s = \int^{t'}_{t}\frac{dt''}{R(t'')} + \Psi(\mathbf{x},
\mathbf{x'}),
\end{equation}
expresses the relation between the retarded time $t'$ and the time
$t$; obtained by setting $s=0$, and the four-current density is
given by
\begin{equation}
J^{m'}(\mathbf{x'}, t') = q \frac{dz}{ds}^{m'} =
q\frac{R(t')v^{m'}}{\sigma}.
\end{equation}
The Lorentz force on a comoving test charge $\tilde{q}$ having
four-velocity $u^{\alpha}$ in the Friedmann models (\ref{FRW1}), is
given by the space part of the four vector
\begin{equation}
f^{\mu} = \tilde{q} F^{\mu\nu}u_{\nu},
\end{equation}
i.e.,
\begin{equation}
\mathbf{F} = -\tilde{q} F^{i0} = \tilde{q} F^{0i}.\label{force}
\end{equation}
In order to obtain expressions for the fields in (\ref{F0k}) for
different Friedmann universes represented by (\ref{f}), consider a
point charge $q$ at the retarded position $(t', \rho')$ moving in
the $\theta=0$ direction such that $\mathbf{v'} =
(v\cos\theta/(R(t'), -v\sin\theta/R(t'), 0)$. To find the components
of the parallel propagator $d_{km'}$ in (\ref{F0k}) consider the
change in the velocity vector as it is parallel transported radially
along a geodesic in 3-spaces of constant curvature given in
(\ref{FRW1}). Then
\begin{equation}
dv^{i} = -\Gamma^{i}_{j1}v^{j}d\rho.
\end{equation}
\subsection{Case k=0}
In this case the radial component of the velocity vector is
unchanged under parallel transport and hence $v^{1} = v^{1'}$ so
that $g_{11'} = 1$. Also $v^{2} = -v\sin\theta/R(t)\rho$ such that
$g_{22'} = \rho$ and $v^{3} = v^{3'} = 0$.  Taking the limit $v <<
c$ the expression for the field in (\ref{F0k}) with $\Psi = \rho$
becomes
\begin{equation}
F_{0k} = \frac{qR^{2}(t')}{\rho R(t)}(g_{km'} +
\rho_{;k}\rho_{;m'})a^{m'} -
\frac{q\rho_{;k}}{R(t)\rho^2},\label{F0kflat}
\end{equation}
where $a^{m'}$ are the components of the acceleration of the point
charge at the retarded coordinates $(t', \rho')$. This expression
reduces to $-\mathbf{E}$ given by (\ref{sciama1}) in the case of
flat spacetime where $R(t) = R(t') = 1$. The total force on the
accelerating point charge $q$ from a charge distribution, can then
be obtained by using (\ref{force}) and integrating over all
contributions. If the distribution is isotropic the $1/\rho^2$ term
in (\ref{F0kflat}) vanishes, and one only needs to consider the
component of the force along the direction of motion, given by
\begin{equation}
F =
\int\tilde{\rho}({x^n})F^{0i}\frac{a_{i}}{|\mathbf{a}|}\,d^{3}x^{n}
=
-\int\tilde{\rho}(x^{n})F_{0i}\frac{a^{i}}{|\mathbf{a}|}\,d^{3}x^{n},
\label{force2}
\end{equation}
where $\tilde{\rho}(x^{n})$ is the charge density of the
distribution and
\begin{equation}
F_{0i}\frac{a^{i}}{|\mathbf{a}|} = \frac{qR(t')}{\rho
R^{2}(t)}a\sin^2\theta.\label{componentflat}
\end{equation}

\subsection{Case k =1}
In this case $v^{1} = v^{1'}$ and $v^{2} =
-v\sin\theta/R(t)\sin\rho$. Therefore $g_{11'} = 1$ and $g_{22'} =
\sin\rho$. Also $\Psi = \rho$ and the expression for the field
(\ref{F0k}) in the limit $v<<c$ is given by
\begin{equation}
F_{0k} = \frac{qR^{2}(t')}{R(t)\sin\rho}(g_{km'} +
\rho_{;k}\rho_{;m'})a^{m'} -
\frac{q\rho_{;k}}{R(t)\sin^2\rho}.\label{F0kclosed}
\end{equation}
The component of the field in the direction of motion is then given
by
\begin{equation}
F_{0i}\frac{a^{i}}{|\mathbf{a}|} =
\frac{qR(t')}{R^{2}(t)\sin\rho}a\sin^2\theta.\label{componentclosed}
\end{equation}

\subsection{Case k =-1}
Again there is no change in the radial component of the velocity
vector under parallel transport and so $g_{11'} = 1$. The parallel
transported angular component is given by $v^2 =
-v\sin\theta/R(t)\sinh\rho$, and so $g_{22'} = \sinh\rho$. The
previous expressions for the field and its component along the
direction of motion are given (for $v << c$) by
\begin{equation}
F_{0k} = \frac{qR^{2}(t')}{R(t)\sinh\rho}(g_{km'} +
\rho_{;k}\rho_{;m'})a^{m'} -
\frac{q\rho_{;k}}{R(t)\sinh^2\rho},\label{F0kopen}
\end{equation}
and
\begin{equation}
F_{0i}\frac{a^{i}}{|\mathbf{a}|} =
\frac{qR(t')}{R^{2}(t)\sinh\rho}a\sin^2\theta. \label{componentopen}
\end{equation}
Sciama's law of inertial induction can be generalized to the curved
spacetime of the Friedmann models by using (\ref{F0kflat}),
(\ref{F0kclosed}) and (\ref{F0kopen}) to represent the gravitational
field of an accelerating particle and (\ref{force2}) for the
gravitational force on this particle from an isotropic matter
distribution instead of the previous expressions in (\ref{sciama1})
and (\ref{sciama2}) which are only valid in flat space. This is done
in the next section.

\section{Inertial Force}

Using the coordinate transformation $r = R_{0}f_{k}(\rho)$, where
$R_{0}$ represents the scale factor at the present time $t = t_{0}$
the Friedmann models in (\ref{FRW1}) can be written in the form
\begin{equation}
ds^2 = -dt^2 + A^2(t)\left[\frac{dr^2}{1 - \kappa r^2} + r^2
d\Omega^2\right], \label{FRW2}
\end{equation}
where $A(t)$ is the dimensionless scale factor $A(t) = R(t)/R_{0}$
and $\kappa = k/R^{2}_{0}$. In these coordinates the expressions for
the component of the field along the direction of motion given by
(\ref{componentflat}), (\ref{componentclosed}) and
(\ref{componentopen}) are given by
\begin{equation}
F_{0i}\frac{a^{i}}{|\mathbf{a}|} =
\frac{qA(t')}{A^{2}(t)r}a\sin^2\theta, \label{componentall}
\end{equation}
for all models. Then using Sciama's law of inertial induction, the
total inertial force on an accelerating particle of mass $m$ in the
Friedmann models is given by
\begin{equation}
F = -\frac{Gma}{c^2}\int\int\int\frac{\tilde{\rho}
A(t')}{r}\sin^2\theta\, \sqrt{g}\,dr\,d\theta\,d\phi,
\label{totalforce}
\end{equation}
where $\tilde{\rho}$ is the density of the universe, and we set
$A(t) = 1$ because the force on the accelerating particle due to
matter at the retarded position $(t', \mathbf{x'})$ is calculated at
the present time $t=t_{0}$. Instead of the radial coordinate in
(\ref{totalforce}) it is more convenient to use the redshift factor
$z = 1/A(t') - 1$ which is related to $r$ by\cite{carroll}
\begin{equation}
r =
\frac{cH_{0}^{-1}}{\sqrt{|\Omega_{c0}|}}f_{k}\left[\sqrt{|\Omega_{c0}|}
\int_{0}^{z}\frac{dz'}{E(z')}\right],\label{r(z)}
\end{equation}
where $f_{k}$ is given by (\ref{f}) and $H_{0}$, $\Omega_{c0}$
represent the Hubble's constant and the current density parameter
associated with spatial curvature respectively. Also for the metric
(\ref{FRW2})
\begin{equation}
\sqrt{g} = \frac{A^3(t')r^2\sin\theta}{\sqrt{1 - \kappa r^2}}; \quad
\kappa = -\frac{H_{0}^2\Omega_{c0}}{c^2},\label{sqrtg}
\end{equation}
and
\begin{equation}
E(z) = \left[\sum_{i(c)}\Omega_{i0}(1+z)^{n_i}\right]^{1/2},
\label{E(z)}
\end{equation}
where $n_{i}$ depends on the source of the density parameter
according to Table \ref{ta1}.
\begin{table}[ht]
\begin{center}
\tbl{Values of $n_{i}$ for different sources in Eq. (\ref{E(z)})}
{\begin{tabular}{lll} \toprule Source & Parameter & $n_{i}$
\\\colrule
matter  & $\Omega_{M}$ & 3  \\
radiation & $\Omega_{R}$ & 4  \\
curvature & $\Omega_{c}$ & 2\\
vacuum & $\Omega_{\Lambda}$ & 0\\\botrule
\end{tabular} \label{ta1}}
\end{center}
\end{table}
Assuming that source of the energy density of the Friedmann models
consists of matter and radiation, the density $\tilde{\rho}$ in
(\ref{totalforce}) can be expressed in terms of the redshift by
\begin{equation}
\tilde{\rho}(z) = \frac{3H_{0}^2}{8\pi G}\left(\Omega_{M0}(1+z)^3 +
\Omega_{R0}(1+z)^4\right). \label{density}
\end{equation}
Substituting (\ref{r(z)})-(\ref{density}) in (\ref{totalforce}) we
get the inertial force on the accelerating particle for the three
Friedmann models
\begin{eqnarray}
%\begin{array}{ll}
k = 0; & F = -ma\displaystyle\int_{0}^{\infty}\frac{\Omega_{M0} +
\Omega_{R0}(1+z)}{(1 +
z)E(z)}\left[\int_{0}^{z}\frac{dz'}{E(z')}\right]\,dz,  \label{k=0}\\[5mm]
k = 1; & F =
-\displaystyle\frac{ma}{\sqrt{|\Omega_{c0}|}}\int_{0}^{\infty}\frac{\Omega_{M0}
+ \Omega_{R0}(1 +
z)}{(1+z)E(z)}\sin\left(\sqrt{|\Omega_{c0}|}\int_{0}^{z}\frac{dz'}{E(z')}\right)\,dz,
\label{k=1}
\\[5mm]
k=-1; & F =
-\displaystyle\frac{ma}{\sqrt{\Omega_{c0}}}\int_{0}^{\infty}\frac{\Omega_{M0}
+ \Omega_{R0}(1 +
z)}{(1+z)E(z)}\sinh\left(\sqrt{\Omega_{c0}}\int_{0}^{z}\frac{dz'}{E(z')}\right)\,dz.
\label{k=-1}
%\end{array}
\end{eqnarray}
Unlike (\ref{berry2}) where the integration is limited to the Hubble
sphere, in our case the integration is extended over the causal
sphere bounded by the particle horizon of the accelerating particle
where $z=\infty$. These includes the inertial contribution from all
matter and radiation in the universe.

\section{Results and Discussion}

In this note we have tried to compute the inertial mass of particles
following the approach of Sciama, i.e. the analogy between
electromagnetic and gravitational forces, assuming that the latter,
just like the former are subject to retardation. As a result, we
assumed that they yield, just like electromagnetism, a far-field
force between material particles of masses $m_1, \, m_2$
proportional to $G m_1 m_2 a/ c^2 r$; note that because of the $1/r$
(rather than the $1/r^2$ appropriate for non-accelerating particles)
force dependence, a spherically symmetric mass distribution yields a
non-zero force at its center.  Based on this analogy we followed
Ref.~\refcite{peters} to compute the electric fields of accelerated
charges in an arbitrary FRW space time and replaced the coupling
constant by $G m_1$ to mimic the effects of gravitational force.
Finally, by integrating this force over all matter in the universe
we computed the gravitational force of the latter on a body of mass
$m$ in relative acceleration (with the universe!). This force is
proportional to the acceleration $a$ with a coefficient that depends
on the cosmological parameters, in (philosophical) agreement with
Mach's principle.

Starting with the spatially flat Friedman model $(k=0)$, the
inertial force on the accelerating particle given by (\ref{k=0}) in
a matter dominated Einstein-de Sitter universe $(\Omega_{M0}=1,\,
\Omega_{\Lambda0} = \Omega_{R0} = 0)$ is given by $F = -(1/3)ma$.
Although this coefficient of $ma$ is larger than that obtained by
Berry, it is still far from unity, and this implies that the total
force on the particle cannot be explained only in terms of the
inertial contribution from all the matter in the universe. In a
radiation dominated universe $(\Omega_{R0} = 1,\, \Omega_{\Lambda0}
= \Omega_{M0} = 0)$ $F = -(1/2) ma$, while if we choose the
currently favored cosmological parameters $\Omega_{M0} = 0.3,
\Omega_{\Lambda0} = 0.7$  we get $F = -0.23\, ma.$

Having said this, we have to point out that the coefficient of $ma$
in all three cases given by (\ref{k=0}) - (\ref{k=-1}) depends on
the \textit{current} values of the cosmological parameters. Since
these parameters depend on the redshift $z$ according to the
expression
\begin{equation}
\Omega_{j} =
\frac{\Omega_{j0}(1+z)^{n_{j}}}{\left[\sum_{i(c)}\Omega_{i0}(1+z)^{n_i}\right]}
\label{omegaj}
\end{equation}
where $n_{i}$ are given in Table \ref{ta1}, this coefficient will be
different at different epochs in the evolution of the universe.
Therefore if $\Omega_{M0} = 0.3,\, \Omega_{\Lambda0} = 0.7$, then at
z = 1 $\Omega_{M1} = 0.77,\, \Omega_{\Lambda1} = 0.23$ and the
inertial force is $F=-0.31 ma $. This higher value of the inertial
force is due to a higher percentage of matter at earlier times. For
very large red-shifts the universe is radiation dominated with
$\Omega_{R} \rightarrow 1$ and $\Omega_{\Lambda} \rightarrow 0$, and
the inertial force thus calculated becomes $F\rightarrow (1/2)ma$.
It is not obvious at this point whether such a variation can lead to
observable effects or inconsistencies with observations; we hope to
be able to look at these issues in the future.

For a spatially closed $(k = 1)$ universe the effect of positive
curvature leads, as expected, to a higher value for the inertial
force on the accelerating particle. So for a slightly curved
universe with $\Omega_{c0} = -0.1,\, \Omega_{M0} = 0.3,\,
\Omega_{\Lambda0} = 0.8$ we get $F = -0.24ma$. As in the flat case
the coefficient of $ma$ is higher in the presence of radiation. So
if $\Omega_{M0} = 0.3,\, \Omega_{\Lambda0} = 0.7,\, \Omega_{c0} =
-0.1,\, \Omega_{R0} = 0.1$, then $F=-0.33 ma$. On the other hand in
the spatially open $(k=-1)$ case, the negative spatial curvature
reduces the inertial force. Taking $\Omega_{M0} = 0.2,\,
\Omega_{\Lambda0} = 0.8,\, \Omega_{c0} = 0.1$, we get $F=-0.19 ma$.
The previous remark about the z-dependence of our results applies
also to the curved Friedmann models.

To summarize, in this article we have used the formalism of
electrodynamics in curved spacetime to generalized Sciama's law of
inertial induction to a curved spacetime represented by the
Friedmann models, and obtained covariant expressions for the
inertial force on a accelerating particle in these models. This
extends previous work by Berry and later by Signore on the subject
of inertia in Friedmann universes. As in Berry's case we have showed
that the total inertial force $F=-ma$ on an accelerating particle
cannot be explained only in terms of inertial induction from the
total matter and radiation in the current universe. We have also
seen from (\ref{omegaj}) that the inertial force is redshift
dependent and increases for earlier times until it reaches the
asymptotic value of $-(1/3) ma$ in the matter dominated case or
$-(1/2) ma $ in the radiation dominated case.

\section*{Acknowledgments}

J.S. gratefully acknowledges financial support from USRA-CRESST and
the University of Malta during his research visit at NASA-GSFC.

%\begin{thebibliography}{000} %for 3 digits

\end{document}